# VM Image Repository and Distribution Models for Federated Clouds: State of the Art, Possible Directions and Open Issues

Nishant Saurabh*, Dragi Kimovski, Simon Ostermann, and Radu Prodan

Distributed and Parallel Systems, Institute of Informatics,
University of Innsbruck, Innsbruck 6020, Austria.
{nishant,dragi,simon,radu}@dps.uibk.ac.at
http://www.dps.uibk.ac.at

**Abstract.** The emerging trend of Federated Cloud models enlist virtualization as a significant concept to offer a large scale distributed Infrastructure as a Service collaborative paradigm to end users. Virtualization leverage Virtual Machines(VM) instantiated from user specific templates labelled as VM Images (VMI). To this extent, the rapid provisioning of VMs with varying user requests ensuring Quality of Service(QoS) across multiple cloud providers largely depends upon the image repository architecture and distribution policies. We discuss the possible state-of-art in VMI storage repository and distribution mechanisms for efficient VM provisioning in federated clouds. In addition, we present and compare various representative systems in this realm. Furthermore, we define a design space, identify current limitations, challenges and open trends for VMI repositories and distribution techniques within federated infrastructure.

## 1 Introduction

The *Cloud Computing* is a ubiqutous global paradigm, empowering users to acquire on demand compute resources without the onus of owning, managing or maintaining them. In this context, one of the important concept is *Infrastructure as a Service* (IaaS)[8] cloud model. Virtualization[9] is a key technology employed in cloud data centers to support IaaS, allowing users to instantiate multiple Virtual Machines (VM). The instantiated VMs constitute users application environment to be adequately scaled by elastic on-demand provisioning in response to variable load to achieve increased utilization efficiency at lower operational cost, while guaranteeing *Quality of Service* (QoS)[7] to end users.

VMs in general, are instantiated using specific templates termed as *VM Images* that are stored in proprietary repositories, leading to provider lock-in[10] and hampering portability or simulataneous usage of multiple federated Clouds. In addition, the proprietary repositories do not take into account underlying application characterstics resulting to deployment and instantiation overheads.

To this end, VMI repository research extends to a novel operational environment aiming to mitigate limitations with regard to VMI storage and distribution for federated cloud infrastructures. Such a Large Scale Distributed Virtual Environment for VMI repository imminently benefit the elastic auto-scaling of diverse applications on cloud resources based on their fluctuating load. Henceforth, VM interoperability across multiple cloud infrastructures is achieved without provider lock-in, only to justify the virtualization technology as a universal cloud IaaS model.

In this paper, we split the state of the art contention into two parts, namely VMI Repository and VMI Distribution. Initially, we emphasize the required consideration to treat image repository beyond the typical storage systems and henceforth, detail the factors defining the VMI Repository with respect to functionality, architecture, VMI management and cloud federation aspects. Furthermore, we discuss the existing VMI distribution tecnhiques and suitability of each with regard to varying VMI repository architecture meant to provide middleware services in federated cloud models.

To examine current advances corresponding to our discussion , we consider as case studies, various production systems, in particular namely: Virtual Machine Repository Catalog (VMRC)[1][11] , Amazon Image Service[2] and Openstack Glance[3]. In our view, these three systems define the closest state of the art of VMI Repository and furthermore, each of the systems has some common and unique set of functionalities to offer. Our discussion focusses on VMI repository service's rationales, distribution models and their respective usage scenario in case of multiple cloud providers. To be concise, we have investigated possible measures required to allow flexibility for rapid VM provisioning appropriated by image repository and distribution models. Finally, we identify open issues and suggest future research directions regarding federated VMI middleware repository.

The contributions of the paper are :

- An overview of the existing storage modelling factors and its application to VM Image repository design.
- An analysis and classification of VM Image storage and distribution techniques applicable to federated cloud models.
- A synopsis of the current state of the research area, identifying trends and open issues.
- A vision on possible future directions.

The remainder of this paper is organized as follows. Initially, Section 2 surveys the existing production systems. Section 3 outlines possible state of the art to design the VM image repository along with the distribution mechanisms in federated cloud infrastructures. We discuss and analyze the quality rationales of surveyed image repository systems in Section 4, followed by possible future directions and open issues. We conclude the paper in Section 5.

---

[1] http://www.grycap.upv.es/vmrc/index.php
[2] http://docs.aws.amazon.com/AWSEC2/latest/UserGuide/AMIs.html
[3] http://docs.openstack.org/developer/glance/

## 2 Existing VMI Repository Systems

The design of most of the existing VMI repositories are subjected to specific hypervisor technology concerning the cloud architecture. In general, users interaction is provided through a web interface and corresponding attached APIs to these repository systems with very basic functionalities. Hence, the VMI and associated metadata management are left onto the user based manual actions instead of allowing automated query executions leading to error-prone usage.

To mention a few such systems, VMware[4] repository system at its disposal allows upload and download of images by authorized users. In addition, a weak virtual management system for existing VMI categorization is intact, enabling to search for the required images corresponding to the target application of the user. Another VMI specific repository by Science Clouds[5] only allow download of exisiting stored images and evades the upload or indexing funtionality of any user specific VMIs. FutureGrid [15], an experimental system for HPC and cloud based applications provisions another image repository with federated storage systems empowering the users to avail upload, download and update funtionality with limited metadata informations through REST interface.

Apart from above mentioned systems, in this section, we give a detailed account of some VMI repositories adopted by private and public cloud infrastructures, namely *VMRC*(Virtual Machine Repository Catalog)[11], *Openstack Glance*[6] and *Amazon Image Service*[7]. In our view, these three systems are closest to the state of the art in the field of VMI repository service with respect to image storage and corresponding image functionalities as a middleware service.

### 2.1 VMRC

The *VMRC*(Virtual Machine Repository Catalogue)[11] modelled as a client-server based architecture enables user to upload, store and catalog VMIs. It also serves as a matchmaking collaborative system to facilitate sharing of images availing through the usage of extensive metadata, where independent users can search and retreive appropriate stored VMI using the catalog funtionality.

In general, VMRC is represented into four modules namely *Storage, Repository, Catalog and Client*. The *Storage* module handles the appropriate mediums to store VMIs, while *Repository* provisions support for transfer of VMIs within different storage mediums. In addition, *Repository* module also facilitates user authorizations in case of VMI uploads. In order to index the stored images, *Catalog* module is used accompanied by unique matchmaking algorithms to retreive the appropriate images suiting the users requirement. The easy usage of the mentioned functionalities are supported with an end-user command line application *Client* module.

---

[4] http://www.vmware.com/appliances
[5] http://scienceclouds.org/marketplace
[6] http://docs.openstack.org/developer/glance/
[7] http://docs.aws.amazon.com/AWSEC2/latest/UserGuide/AMIs.html

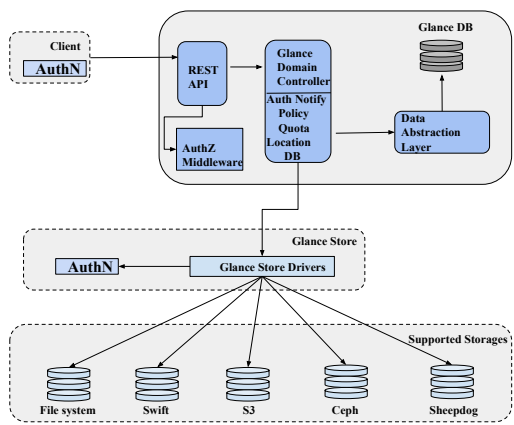

Fig. 1: Glance Architecture

### 2.2 OpenStack Glance

*Glance*[8] in general, is a middleware service enabling users to upload independent data assets including VMI. In particular, glance image service provisions various functionalities including discovering, registering and retrieving images. In order to provide respository based service, federation of storage systems are attached. These storage systems with varying capabilities ranging from simple file systems to object based storage are located within varying regions to manage VMI services.

Glance integrated with Openstack virtualized infrastructure, follows a client-server based centralized architecture which provides a REST API for its users to access image functionality. Furthermore, it provides an interface to its various components managing internal operations as shown in figure 1 to openstack. Any REST API based request from the client is accessed through *domain controller* component which handles services corresponding to different layers, where each layer appropriates to perform a specific task. These tasks include authorization governing policies regarding the actions of a user to a particuar image such as verifying access rights to add, update or delete a VMI or checking quota of storage capacity attached to a user for adding an image at a particular region etc. It is to be observed that policies regarding the authorization, storage quota could vary and depend upon the organization implementing glance domain controller component specific to its infrastructure.

Another component *Glance Store*, handling VMI storage provides an uniform access to various attached storage systems. It provides a series of library functions to execute VMI operations requested by the user with regards to au-

---
[8] http://docs.openstack.org/developer/glance/

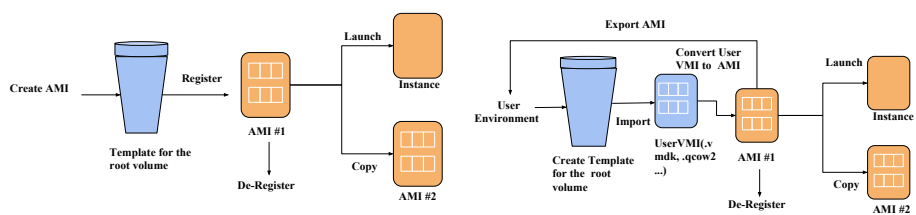

(a) Life Cycle of AMI creation onto AWS.

(b) Life cycle of Upload of User VMI onto AWS

Fig. 2: AWS VM Image Services depicting creation and Upload of amazon infrastructure supported User VMI.

thorization inputs received from *Domain Controller*. The library functions are basically file based operations such as upload, update , delete etc.

The *Domain Controller* also provides an interface to centralized *Glance Database* API, which contains several methods for moving image metadata to and from attached persistent storage systems. These methods basically references to metadata regardging creating , updating, retreiving VMI with respect to parameters like image identifier, image location, image context etc. Once image is registered onto the centralized database, it is deemed appropriate to be instantiated with specific configurations within a particular region or loaction.

### 2.3 Amazon Image Service

*Amazon Elastic Compute Cloud*(EC2) services is one of the most poular commerical public cloud infrastructure. In the early stages, *Amazon Web Services*(AWS) only provided functionality to create *Amazon Machine Images*(AMI)[9] onto its own infrastructure instead of allowing upload of user specific images as shown in figure 2a. The AMI is similar to VMI, which includes a template for the root volume for VM to be instantiated consisting of OS, application server and underlying target application services. Furthermore, AMI also comprise of permisssion authorizations to launch corresponding AMI.

However, recently AWS added a VM export/import functionality to import and export VMI from user specific environment onto Amazon EC2. This functionality enables a user to include its own configurations, security and compliance requirement within image intended for target VM instance.

AWS provides a client interface to upload VMI. As a part of import functionality, user specific images are converted to AWS EC2 AMI and stored onto

---

[9] http://docs.aws.amazon.com/AWSEC2/latest/UserGuide/AMIs.html

*Elastic Block store* or *S3* data store of Amazon. The AMI identifier is further mapped onto a region as specified by the user, hence facilitating the instatiation of VMs. AWS also allows the user to have the authority of enabling the stored images to be either private, shared with specific AWS users or to be public to whole community.

## 3  State of the Art

In Section 2, we overviewed the existing image repository systems. Although most of them support the basic functionalities of upload and download of VMIs, the eminent federation functionalities including repository management, interoperability etc are left onto the user based manual actions. In this section, we focus on the possible state of the art in terms of VMI storage and distribution for federated clouds. While we state some of the common functionalities, we also define the VMI and corresponding repository operations, currently missing in the existing production systems.

### 3.1  VM Image Storage Repository

VM Image usually in size of GigaByte(GB) contains a bare operating system (OS), or an operating system with user defined software and applications. In specific cases, additional data is also attached to corresponding image template. A typical example of such VMI is running an Earth Observational Data processing cloud application[12] with large sized sattelite imagery data. The storage of such diverse images require a scalable and elastic storage model for optimized VMI distribution across the multiple cloud providers. Furthermore, image repository is also necessitated to act as a middleware providing services beyond the typical storage repositories. Placek et al.[6] defines storage systems taxonomy built upon a number of factors. To this extent, we discuss the state of the art of VMI storage Repository including functionality, architecture and federated interoperability concerning VMI application characteristics.

**Functionality** The VMI Repository is customary to have a wide array of behavioral functions beyond the typical storage and retreival offered by general purpose storage systems. In custom, a VMI is a collection of complex set of bytes with a sequence of functional descriptions specific to user defined applications. A large sized VMI can also be splitted into fragments, where each fragment refers to a specific funtionality[13]. These attributes make it difficult to inject updates if any, directly to the stored VMI or functional fragments. In case of decentralized repository with geographically distributed storage systems, propagating updates to various stored image replicas is even more of a gruelling task. Hence, VMI repository is specifically to be characterised as a persistent storage system restricted to the write-once and read-many feature. In this category of storage, any updates to the VMI propells the removal of old image and creating a new image onto the repository. However, the concept of VM contextualization can be utilized by the incorporation of existing tools like *Chef*[10] and *Puppet*[11], hence facilitating VMI size optimization by synthesizing and pruning the un-necessary files or analyzing and fragmenting the VMI with respect to functionality in case of large sized images encompassed with various services. This feature indeed helps in reducing the storage cost with faster distribution and VM provisioning across cloud boundaries.

In addition, the modern cloud providers maintains a list of VMI's provisioned for sharing amongst users. These images are typically not user specific, instead consists of some most commonly used OS platforms or applications. The repositories facilitating sharing of such images deliver pubish/share functional service, inhibiting the censorship of stored VMI.

One of the other interesting functionality of VMI repository is providing a homogenous interface to an array of attached storage systems. These evident systems either coupled or decoupled to cloud storage are accomplished with varying capabilities which provide unique interface to interact with. In such cases VMI Repository has much of a task to act as a middleware entity instead of just a storage service.

**Repository Architecture** The repository architecture in general, determines the operational boundaries of stored resource, ultimately forging behavior and functionality corresponding to the application services, a resource provide[6]. In our paper, VMI is the stored resource and the operational boundary corresponds to the factors affecting distribution of images to multiple cloud providers. Typically, image storage repository can be classified as *Centralized* or *Distributed* on the basis of the architecture it follows. In this section, we discuss the functional capabilities and limitation of pre-mentioned architectural models to the applications of VMI.

*Centralized.* In most of cloud infrastructures, a centralized image server serves as a repository to host a catalog of VMIs. These repositories maintain a central index of stored images which are either produced locally or imported from user specific environment.

In general, Centralized repository can be either classified as globally or locally central[6]. The globally centralized model contains a single image server handling requests for many users related to VMI functionality such as upload, update, download etc. Such architecture has limited scalability with a single point of failure.

The image repositories within cloud data centres broadly come into category of locally centralized architectures which alleviate independent functionalities across multiple attached servers. However, VMI repositories under this category as well, faces scalability bottlenecks and failure centric issues, specifically in the case of supporting federated cloud models, where each provider regulate its own trust policies.

---

[10] http://www.opscode.com/chef
[11] http://www.puppetlabs.com

*Distributed.* The recent advances in storage repository architecture has observed existing centralized models evolving into decentralized approaches to achieve scalability and reliability. The reason being, centralized structured models often encounter bandwidth and scalability bottleneck, hence influencing the quality of service.

The essential feature of distributed repository is to compound the image stores within multiple cloud providers interfaced with independent APIs, to be precise a middleware service providing user transparency for the VMI storage at different attached storage systems. Another essential characterstic is to maintain the VMI replicas or chunks placement with respect to fault tolerance techniques used such as *Replication* and *Erasure Coding*[14] respectively in consireation to reduce distribution times aross cloud sites.

**VMI Repository Management** The distributed VMI repository enables to maintain a set of VMI replicas or erasure coded chunks to enhance fault tolerance. However, it is as imminent to decide the repository nodes at which replicas should be placed. Initially, the user provides a set of metrics including storage cost, performance based metrics while uploading the image. Moreover, the attached storage systems are accomplished with varying capabilities, hence exists different cost policies and performance metrics for each. The VMI repository system applies a decision making process, placing the replicas onto the storage repositories satisfying the user specifications for initial upload.

Furthermore, every time a user requests for distribution of image to a cloud provider, a learner module track the statistics of the frequency of distribution of image to a specific provider. To this extent, the placement of VMI replicas or chunks concerning factors like image popularity at a particular cloud provider or across cloud boundaries, avoiding vendor lock-in etc is reshuffled to the image storage repository closer to the region corresponding to the provider with frequent distributions. This greatly improves the geographical scalability of stored images with respect to faster distribution and provisioning.

**Federation** VM Images are currently stored by cloud providers in proprietary centralized repositories without considering application characterstics and their runtime requirements, causing high deployment and instantiation overheads. Moreover, users are expected to manually manage the VM Image storage which is tedious, error-prone and time-consuming especially if working with multiple cloud providers. Every cloud provider is highly interested in attracting new customers from other providers. Unfortunately, current users must be familiar with providers repository interfaces and specific VMI formats in order to use them, which is unsurpassable barrier in deploying new images and exploiting provider resources.

The VMI repository for federated cloud models mitigate the user limitations and manages the interoperability of user created images across multiple providers. Once a request is received by the repository to distribute a corresponding VMI onto a cloud provider, an image conversion module is executed to convert VMI to the format suited for the cloud infrastructure, it has to be instantiated on. Hence, facilitating the user with a federation middleware VMI repository, servicing storage and distribution requests of images across a federation of cloud providers to achieve globalised Infrastructure as a Service paradigm.

### 3.2 VMI Distribution

Modern cloud computing data centers face the key challenge to provide rapid VM provisioning in elastic and scalable manner. To this extent efficient VMI distribution[1],[2],[3],[4],[5] onto the physical compute node across cloud providers is an imminent aspect. The distribution process essentially suffers a handicap in case of federated cloud models owing to the inconsiderate VMI Repository architecture offering unscalable services to increasing user requests, and lack of VMI interoperability across multiple clouds as discussed earlier. In this section we discuss some of the popular VMI distrbution techniques, focussing to its appropriateness and limitation with reference to repository models for federated clouds.

### 3.3 Unicast Distribution

Unicast distribution[3], a fairly simple method for distributing VMI works for centralized as well as decentralized image repositories. The VMIs of appropriate format are transferred from the image repository to the destined cloud provider in a sequential manner. This method has a huge drawback in terms of transfer rate specific to increased number of requests within a time interval.

**Binary Tree Distribution** In contrast to the naive sequential approach used by Unicast Distribution, binary tree based distribution[3] model follows the parallelized transfer of images. The technique arranges the compute nodes as balanced binary tree. The parent node initiates the image transfer in a sequential fashion followed by the transmission from child nodes at respective levels. However, the transfers are synchronized at every level of the tree to avoid the initiation of transmission from child node until parent's node data is available. Once the intial image transfer from the parent node completes, the receiving node becomes parent itself.

Binary tree distribution of images optimizes the throughput at a lower distribution rate. This technique suits the distrbuted VMI repository architecture, however application within a cross cloud environment is an area of concern with regard to trust policies between multiple infrastructures.

**Multicast Distribution** The multicast distribution[3] technique is mostly preferred in local environment. The image chunk packets are distributed to compute nodes registered onto the host node subscribed for multicast transfer. However multicasting of image is not preferred in case of transferring data over network boundaries specifically in the case of multple cloud providers requiring special multicast protocol support at the core of their internal network.

**Peer-to-Peer Distribution** In case of Peer-to-Peer distribution[3], a popular bit-torrent protocol[4] is used to distribute VMI to corresponding compute nodes. Using this technique, a torrent file is generated comprising of the URL of the tracker node storing the VMI. Furthermore, the storage node executes the seeder module, to which bit-torrent client started on specific compute nodes across multiple cloud providers interface with. To this end, the compute nodes connect to the tracker using URL and seed images from the host storage node completing efficient transmission.

## 4  Discussion

In this section, we summarize the main features of three Image repository systems surveyed in Section 2. We lead our discussion further by focusing on system-wise decision rationales and possible future research directions.

### 4.1  Summary

In terms of typical storage systems, the systems we overviewed does provide basic funtionalities including upload, store and update VMI. On one hand, VMRC provisions indexing of images via *Catalog* functionality, while Amazon allows publish/share of VMIs with respect to appropriate authorization in each case. Although, the discussed production systems qualify for the VMI storage functionality, none of them provide service to facilitate interoperability of images over multiple cloud providers. As mentioned, Openstack Glance and Amazon comprise of proprietary image repository, while VMRC doesnt contribute to interoperability issue, instead has a unique VM matchmaking service for sharing of images. Moreover, the locally centralized architectural model of defined systems inhibit scalable image distribution and hence amounts to delayed VM provisioning. Specifically, the current state of the art in consideration with these respective systems represents a wide gap compared to the possible state of the art for VMI Repository and Distribution models for federated clouds.

### 4.2  Possible Directions and Open Issues

Based on the survey of studied systems and possible state of the art presented in the paper, we propose visions on directions and open issues. One of the promising orientation in this domain, in our view, is interoperability and portability support of VMIs over multiple providers by image repositories. This is particularly important to realise the Cloud IaaS as an all-inclusive paradigm. One way of enhancing interoperability lies in the managment of images by introducing the vendor lock-in objective in consideration to trade-off establishment with $QoS$ cloud metrics and providing a set of optimal solutions to the user with image store options to avoid vendor lock-in. This would require extensive analysis of metadata informations of specific VMIs including funtional descriptions and requirements. An another way of solving interoperability lies in VM contextualization, where VMIs stored functional fragments can be assembled by minimal virtual machines running at destination cloud sites with specific requirements.

Secondly, VMI repository is required to enforce optimization techniques for VMI replica management over the distributed repository to enhance the distribution of images for rapid VM provisioning. In particular, the distribution techniques and its application in different cloud environments, to be precise within same and cross cloud networks is needed to be included as an optimization objective.

## 5  Conclusion

VMI Repository systems and distribution mechanisms attibuted to underlying VMI characteristics is a promising and essential research area. However, there is a need to look beyond the typical storage systems with regard to VMI operational boundaries in terms of efficient distribution and VM provisioning. Henceforth, realizing IaaS as a cloud service beyond a specific provider. In this regard, we discussed the possible state of the art in VMI Repository and Distribution models. We pointed out various factors to define a design space for image repository and prior contributing scenarios to federated infrastructure. We also compared three representative image repository systems identiying the existing gap between current state of the art and the possible design space. Hence, highlighting some of the open issues and possible future directions, including VMI management as a repository service for enhanced distribution, image interoperability support across multiple providers.

### Acknowledgments.

This work was accomplished as a part of project *ENTICE: "dEcentralised repositories for traNsparent and efficienT vIrtual maChine opErations"*[12], funded by the European Unions Horizon 2020 research and innovation programme under grant agreement No 644179. The authors would also like to thank anonymous reviewers for their valuable comments.

---

[12] http://www.entice-project.eu/